 \documentclass[preprint,prc,showpacs,preprintnumbers,
                 superscriptaddress,amsmath,amssymb,floatfix]{revtex4}

\usepackage{graphicx}
\usepackage{dcolumn}
\usepackage{bm}
\usepackage{amsmath}
\usepackage{amsfonts}
\usepackage{amssymb}
\usepackage[]{hyperref}

\begin{document}
\title{Three-dimensional angular momentum projected relativistic point-coupling
 approach for the low-lying excited states in $^{24}$Mg}

 \author{J. M. Yao}
 \email{jmyao@pku.edu.cn}
 \address{School of Physics and State Key Laboratory of Nuclear Physics and
 Technology, Peking University, 100871 Beijing, China}
 \address{Physik-Department der Technischen Universit\"at M\"unchen, D-85748
 Garching, Germany}
 \author{J. Meng}
 \email{mengj@pku.edu.cn}
 \address{School of Physics and State Key Laboratory of Nuclear Physics and
              Technology, Peking University, 100871 Beijing, China}
 \address{Institute of Theoretical Physics, Chinese Academy of
          Sciences, Beijing, China}
 \address{Center of Theoretical Nuclear Physics, National
          Laboratory of Heavy Ion Accelerator, 730000 Lanzhou, China}
 \author{D.~Pena~Arteaga }
 \address{Physik-Department der Technischen Universit\"at M\"unchen, D-85748
         Garching, Germany}
 \author{P. Ring}
 \address{Physik-Department der Technischen Universit\"at M\"unchen, D-85748
         Garching, Germany}
\date{\today}

\begin{abstract}
 A full three-dimensional angular momentum projection on top of a
triaxial relativistic mean-field calculation has been implemented
for the first time. The underlying Lagrangian is a point coupling
model and pairing correlations are taken into account by a monopole
force. This method is applied for the low-lying excited states in
$^{24}$Mg. Good agreement with the experimental data is found for
the ground state properties. A minimum in the potential energy
surface for the $2^+_1$ state, with
$\beta\simeq0.55,\gamma\simeq10^\circ$, is used as the basis to
investigate the rotational energy spectrum as well as the
corresponding B(E2) transition probabilities as compared to the
available data.
\end{abstract}

\maketitle

One of the most discussed topics since the 60's has been the
possible triaxial nature of 2s-1d shell nuclei. The best candidate
to show such a deformation is the nucleus $^{24}$Mg, which has been
extensively studied both experimentally and with state-of-the-art
theoretical models. However, the question of whether it is triaxial
is still not well settled.

Early experiments and their interpretation, such as the
spectroscopic factors extracted from the $^{24}$Mg(d,p)$^{25}$Mg
reaction data done by Parikh~\cite{Parikh68}, favor a prolate axial
deformation. On the other hand, there is also evidence that supports
the triaxiality of $^{24}$Mg, like the interpretation of inelastic
polarized proton scattering by Lombard et al.~\cite{Lombard70}, or
the existence of a low-lying $K=2$ band~\cite{Endt78}. However, the
interpretation of this band as a rotational band or a band built on
one-phonon $\gamma$-vibrational states, is the origin of much
controversy. Phenomenologically, the energy spectrum for the
low-lying states of $^{24}$Mg is well described by the asymmetric
rotor model of Davydov-Filipov with a triaxiality parameter
$\gamma=22^\circ$, while the branching ratios and the reduced
transition rates are not~\cite{Meyer72}. Furthermore, the interband
transitions are reproduced with $\gamma=14^{\circ}$, a value that
cannot give the correct energy spectrum~\cite{Branford75}.

Many microscopic studies have been done to describe the shape of
$^{24}$Mg. So far, most of them have relied on investigations of the
potential energy surface (PES) within various mean-field
approaches~\cite{Ring80}. Early triaxial Hartree-Fock (HF)
calculations with different Skyrme forces showed that the shape
depends strongly on the spin-orbit part of the
interaction~\cite{Grammaticos75}. With a proper treatment of this
term, both Hartree-Fock-Bogoliubov (HFB) calculations with the Gogny
force~\cite{Girod83} as well as HF+BCS investigations with the
Skyrme force~\cite{Bonche87}, yield axially symmetric prolate
minima. However, a Three Dimensional (3D) cranked HFB calculation
within the pairing-plus-quadrupole model predicts a substantial
triaxial deformation for states with higher angular momenta $J>4$ in
the ground state band~\cite{Oi05}. Similar investigations have been
also carried out within the relativistic mean-field (RMF) approach,
and an axial prolate ground state has been predicted for $^{24}$Mg
with either NL1~\cite{Koepf88} or TM1~\cite{Hirata96} parameter
sets.

All the mentioned microscopic studies are based on the mean-field
approximation, which is able to take into account correlations by
``spontaneous symmetry breaking'', e.g., by the introduction of a
deformed mean-field that breaks rotational symmetry in the intrinsic
frame. In order to compare properly with the experimental data, one
has to transform to the laboratory system and to resort to Angular
Momentum Projection (AMP) to restore the broken symmetry. However,
due to its very demanding numerical nature, only very recently it
has been possible to apply such a projection procedure to study the
low-lying excited states of $^{24}$Mg, in the context of the
Generator Coordinate Method (GCM), for example with the Skyrme SLy4
force~\cite{Valor00}, the Gogny D1S force~\cite{Guzman02} and the
relativistic point-coupling model with PC-F1 force~\cite{Niksic06}.
These investigations have shown that the energy gain due to the
restoration of rotational symmetry is of the order of several MeV.
Meanwhile, it has to be pointed out that all these AMP+GCM studies
are restricted to axially symmetric mean-fields. Although great
success was achieved in the description of both the spectrum and
B(E2) transition probabilities with $J<6$, neither the triaxial
deformation information nor a sa\-tis\-factory description of the
observed $K=2$ band could be obtained. Therefore, it is essential to
introduce triaxial quadrupole deformation at the mean-field level
and perform full three-dimensional angular momentum projection
(3DAMP).

In the context of the phenomenological models, 3DAMP has been done
using small shell model spaces and the corresponding effective
interactions~\cite{Hara82,Hayashi84,Burzynski95,Enami99}. It is
shown that the restoration of rotational symmetry has a strong
influence on the topological structure of ($\beta, \gamma$) energy
surface~\cite{Hayashi84}. In the context of energy density
functionals, 3DAMP has been performed on top of Hartree-Fock with a
simple Skyrme-type interaction~\cite{Baye84}, or with a full Skyrme
energy functional~\cite{Zdunczuk07}. In both cases cranked wave
functions were projected to approximate a variation after projection
procedure, but pairing correlations were not included. Only very
recently, projection on particle number (PNP) and angular momentum
with configuration mixing has been attempted in a triaxial HFB
theory for the study the low-lying excited states of
$^{24}$Mg~\cite{Bender08}.

RMF theory, which relies on the ideas of effective field theory and
of density functional theory, represents a very successful approach
to low energy nuclear structure studies with a few universal
parameters~\cite{Serot86,Ring96,Vretenar05,Meng06}. It incorporates
many important relativistic effects from the beginning, such as the
presence of large Lorentz scalar and vector mean fields with
approximately equal magnitude and opposite sign that leads to a new
saturation mechanism via the difference between the scalar and
vector densities, and yields naturally the large spin-orbit
splitting needed for the understanding of magic numbers in finite
nuclei. In this manuscript, for the first time, 3DAMP has been
implemented on top of triaxial relativistic mean-field theory based
on point-coupling (RMF-PC) interactions, and will be applied in this
Letter to study the low-lying excited states of $^{24}$Mg.

In this approach, the intrinsic mean-field state is obtained through
the minimization of an energy functional based on the following
Lagrangian density~\cite{Burvenich02}
 \begin{eqnarray}
  \label{Lagrangian}
 {\cal L}&=&\bar\psi(i\gamma_\mu\partial^\mu-m)\psi\nonumber\\
         &&
             -\frac{1}{2}\alpha_S(\bar\psi\psi)(\bar\psi\psi)
             -\frac{1}{2}\alpha_{V}(\bar\psi\gamma_\mu\psi)(\bar\psi\gamma^\mu\psi)\nonumber\\
         &&
             -\frac{1}{2}\alpha_{TS}(\bar\psi\vec\tau\psi)\cdot(\bar\psi\vec\tau\psi)
             -\frac{1}{2}\alpha_{TV}(\bar\psi\vec\tau\gamma_\mu\psi)\cdot(\bar\psi\vec\tau\gamma^\mu\psi)\nonumber\\
         &&
             -\frac{1}{3}\beta_S(\bar\psi\psi)^3-\frac{1}{4}\gamma_S(\bar\psi\psi)^4
             -\frac{1}{4}\gamma_V[(\bar\psi\gamma_\mu\psi)(\bar\psi\gamma^\mu\psi)]^2\nonumber\\
         &&  -\frac{1}{2}\delta_S\partial_\nu(\bar\psi\psi)\partial^\nu(\bar\psi\psi)
             -\frac{1}{2}\delta_V\partial_\nu(\bar\psi\gamma_\mu\psi)\partial^\nu(\bar\psi\gamma^\mu\psi)\nonumber\\
         && -\frac{1}{2}\delta_{TS}\partial_\nu(\bar\psi\vec\tau\psi)\cdot\partial^\nu(\bar\psi\vec\tau\psi)\nonumber\\
         &&
             -\frac{1}{2}\delta_{TV}\partial_\nu(\bar\psi\vec\tau\gamma_\mu\psi)\cdot\partial^\nu(\bar\psi\vec\tau\gamma^\mu\psi)\nonumber\\
         &&
             -\frac{1}{4}F^{\mu\nu}F_{\mu\nu}-e\frac{1-\tau_3}{2}\bar\psi\gamma^\mu\psi A_\mu .
 \end{eqnarray}
  which contains eleven coupling constants, $\alpha_S$,
$\alpha_V$, $\alpha_{TS}$, $\alpha_{TV}$, $\beta_S$, $\gamma_S$,
$\gamma_V$, $\delta_S$, $\delta_V$, $\delta_{TS}$ and $\delta_{TV}$,
where the subscript indicates the symmetry of the coupling: $S$
stands for scalar, $V$ for vector and $T$ for isovector, while the
Greek letters indicate the kind of contact interaction: $\alpha$
refers to four-fermion terms, $\beta$ and $\gamma$ to third- and
fourth-order terms respectively, and $\delta$ to derivative
couplings.

From the Lagrangian density in Eq.(\ref{Lagrangian}), one can easily
obtain an energy density functional, whose minimization provides the
intrinsic mean-field state wave-function. So far, pairing
correlations have been taken into account within the BCS approach
based on a monopole force and a smooth cutoff factor is taken into
account to simulate the effects of finite range~\cite{Bender00}. To
obtain the PES, the mass quadrupole moments are constrained through
the quantities $q_{20}$ and $q_{22}$, which are related to the
triaxial deformation parameters $\beta$ and $\gamma$ of the Bohr
Hamiltonian by $q_{20}=\frac{3A}{4\pi}R^2_0\beta\cos \gamma$ and
$q_{22}=\frac{3A}{4\pi}R^2_0 \frac{1}{\sqrt{2}}\beta\sin\gamma$,
where $R_0=1.2A^{1/3}$ fm. The total mass quadrupole moment $q$ is
thus given by~$q=\sqrt{16\pi/5}\sqrt{q^2_{20}+2q^2_{22}}$. During
minimization, parity, $D_{2}$ symmetry, and time-reversal symmetry
are imposed. The densities are thus symmetric with respect to
reflections on the $x=0$, $y=0$ and $z=0$ planes. The parameter set
chosen for the Lagrangian density in Eq.(\ref{Lagrangian}) is
PC-F1~\cite{Burvenich02}. The solution of the equations of motion
obtained from the energy density functional is accomplished by
expansion on a set of isotropic three-dimensional harmonic
oscillator basis in Cartesian coordinates with $N_{\rm sh}=8$ major
shells. The oscillator length is chosen to be $b_{0} = \sqrt{
\hbar/m\omega_{0} }$ with $\hbar\omega_0=41 A^{1/3}$ MeV.  The
pairing strength constants $G_n=34.60/A$ and $G_p=33.75/A$ are
determined separately for neutrons and protons by adjusting the
pairing gaps at the minimum of mean-field PES to the odd-even mass
difference as obtained with a five-point formula, and kept fixed
throughout the constraint calculation.

The wavefunction $\vert \Psi^{JM}_{\alpha,q}\rangle$ with good
quantum numbers $\hat{J}$ and $\hat{J}_{z}$ has the form,
 \begin{equation}
    \label{TrialWF}
    \vert \Psi^{JM}_{\alpha,q}\rangle
    = \sum_{K\geq0} f^{JK}_{\alpha}(q) \vert JMK+,q\rangle
 \end{equation}
where the sum is restricted to non-negative even values of $K$. The
index $\alpha=1,2,\cdots$ labels the different collective
excitations. The angular momentum projected $K$-component, $\vert
JMK+,q\rangle$, is given by
 \begin{equation}
    \vert JMK+,q\rangle
    =\dfrac{1}{1+\delta_{K0}}[\hat P^J_{MK}+(-1)^J\hat P^J_{M-K}]\vert\Phi(q)\rangle.
 \end{equation}
 where the projection operator $\hat P^J_{MK}$ is defined as~\cite{Ring80}
 \begin{equation}
    \hat P^J_{MK}=\frac{2J+1}{8\pi^2}\int d\Omega D^{J\ast}_{MK}(\Omega)
    \hat R(\Omega),
 \end{equation}
 with $\Omega$ representing the set of three Euler angles.
$D^{J}_{MK}(\Omega)$ is the corresponding Wigner function. The
expansion coefficients $f^{JK}_{\alpha}(q)$ are determined by the
generalized eigenvalue equation,
 \begin{equation}
    \label{AMPWHE}
    \sum_{K^\prime\geq0}\{{\cal H}^J_{KK^\prime}(q;q)
    - E^J_\alpha{\cal N}^J_{KK^\prime}(q;q)\}f^{JK^\prime}_{\alpha}(q)=0,
 \end{equation}
where the overlap kernels (${\cal O}={\cal N}, {\cal H}$) are given
by:
 \begin{equation}
    \label{OverlapK}
    {\cal O}^J_{KK^\prime}(q;q)
    =\langle JMK+q\vert \hat O\vert JMK^\prime+q\rangle,\quad \hat O=1, \hat H.
 \end{equation}
The overlap kernels are determined with the help of generalized
Wick's theorem~\cite{Onishi66,Balian69}. In addition the method of
Neergard and W\"{u}st~\cite{Neergard83} is also used to determine
the phase of the overlaps $\langle\hat R(\Omega)\rangle$, and it is
found that both methods are in good agreement. The norm overlaps for
intrinsic mean-field states with deformation $\gamma=0$ have been
compared with those computed in the Gaussian overlap approximation,
which turns out to be a good approximation, as already pointed out
in Ref.~\cite{Niksic06}. This provides a very useful test of the
numerical implementation of the angular momentum projection
techniques.  For the Hamiltonian overlap $\langle\hat H\hat
R(\Omega)\rangle$, transition (or mixed) density matrices are
adopted.  The numerical evaluation of the kernels (\ref{OverlapK})
is carried out by a N-point Gauss-Legendre quadrature over the Euler
angles $\Omega=(\phi,\theta,\psi)$. With $N_\theta=14,
N_\phi=N_\psi=24$, it is possible to achieve a precision of
$0.001\%$ in the energy of a projected state with an angular
momentum up to $J=6$. With the consideration of symmetries in the
overlaps~\cite{Hara82,Enami99}, only 1/16 of the total overlaps
needs to be evaluated directly.

The standard procedure described in Ref.~\cite{Ring80} has been used
to solve the generalized eigenvalue equation (\ref{AMPWHE}). By
construction, the collective wave function Eq.(\ref{TrialWF}) avoids
the appearance of zero eigenvalues, and therefore there are $J/2+1$
or $(J-1)/2$ collective states and rotational energy levels for even
or odd angular momentum $J$~\cite{Enami99}, respectively. These
levels can be classified into different bands according to their
B(E2) transition probabilities,
 \begin{equation}
    B(E2; q,J_i,\alpha_i\rightarrow q,J_f,\alpha_f)
    = \frac{e^2}{2J_i+1}
    \vert\langle J_f,q\vert\vert \hat Q_{2}\vert\vert J_i,q\rangle \vert^2.
 \end{equation}
  Since these quantities are calculated in full configuration
space, there is no need to introduce effective charges, and hence
$e$ denotes the bare value of proton charge. A full detailed
discussion about the 3DAMP+RMF-PC approach can be found in
Ref.~\cite{Yao08}.

\begin{figure}[h!]
 \centering
 \includegraphics[width=5cm]{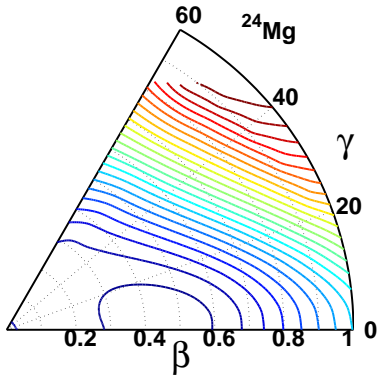}
 \caption{(Color online) The potential energy
 surface in $\beta$-$\gamma$ plane obtained
 by triaxial RMF-PC+BCS calculation for $^{24}$Mg.
 The contour lines are separated by $1.0$~MeV.}
 \label{fig1}
\end{figure}

 Figure~\ref{fig1} shows the PES of $^{24}$Mg in the ($\beta,
 \gamma$) deformation plane, calculated with triaxial RMF-PC+BCS. The
 minimum with $E=-193.57$~MeV, and $\beta_p=0.48$ is found which is less bound
 while compared with the data $E=-198.26$~MeV~\cite{Audi03}.

\begin{figure}[]
 \centering
 \includegraphics[width=4cm]{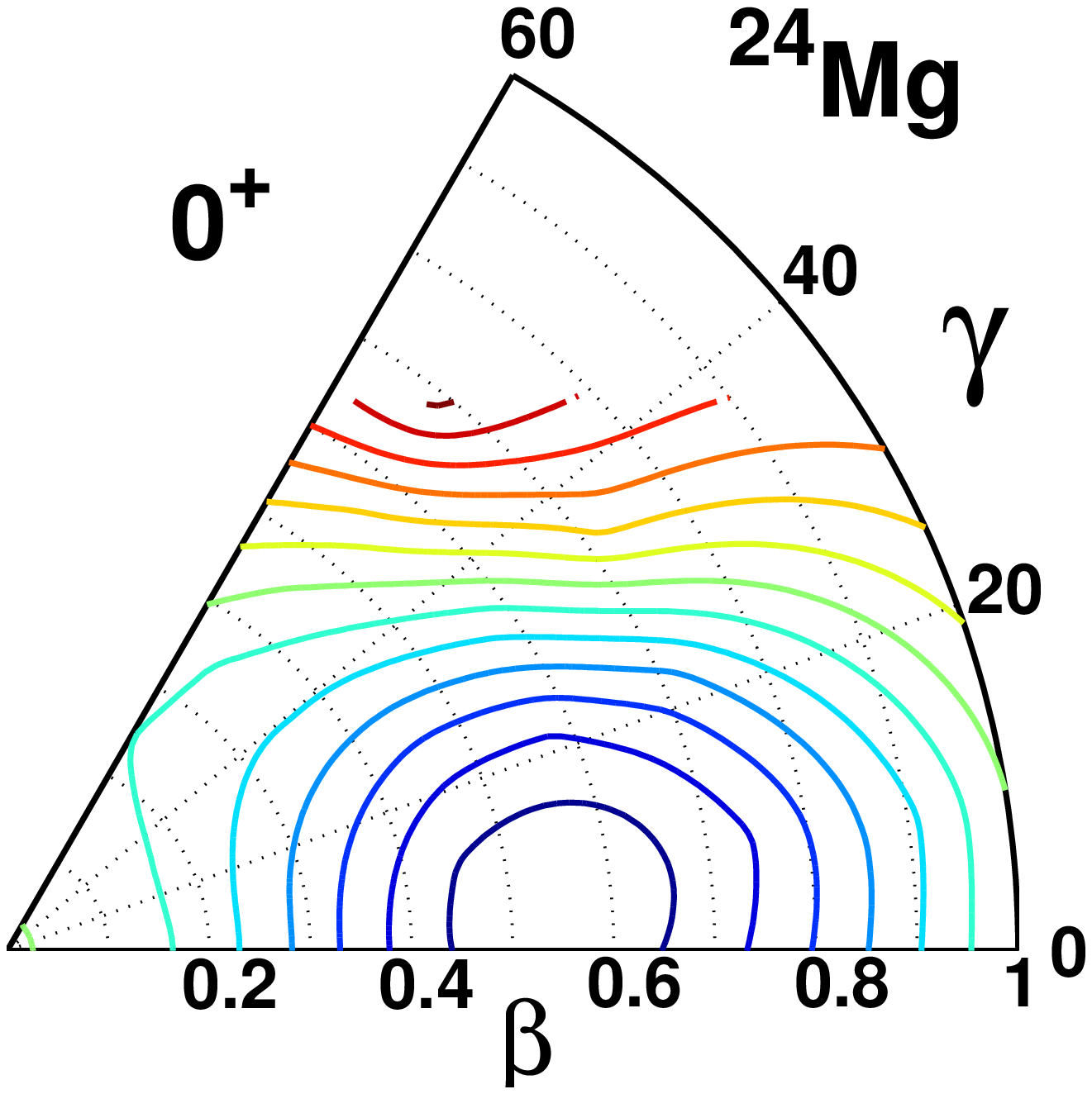}
 \includegraphics[width=4cm]{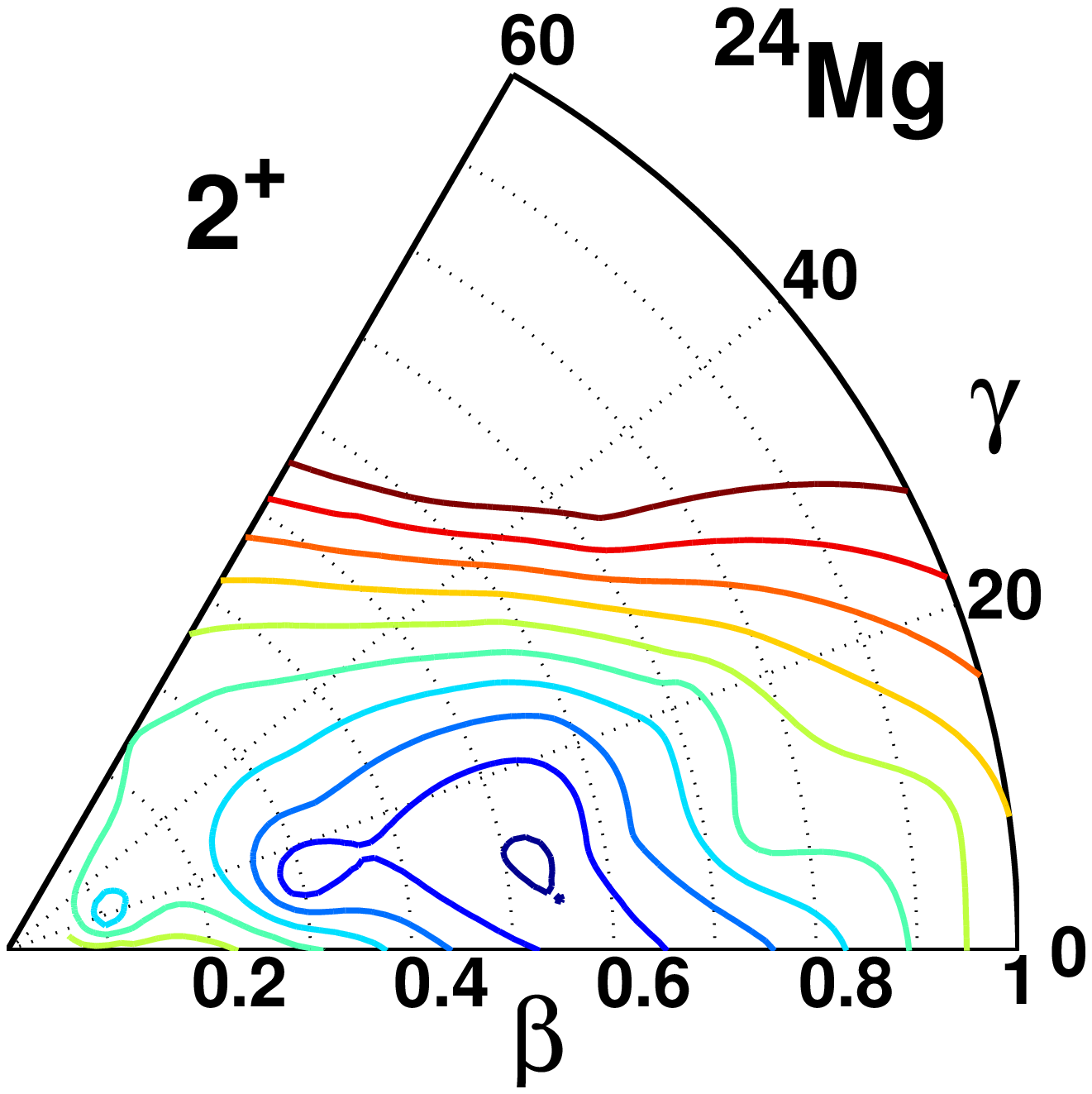}
 \caption{(Color online) Three-dimensional angular momentum projected
 energy surfaces with $J=0$ (left panel)
 and $J=2$ (right panel) in the $\beta$-$\gamma$ plane for
 $^{24}$Mg. The contour lines are separated by $1.0$~MeV}
 \label{fig2}
\end{figure}

The PES of the first $J^\pi=2^+$ and $J^\pi=0^+$ projected states
for $^{24}$Mg in the $\beta$-$\gamma$ plane are plotted in
Fig.~\ref{fig2}. There is no pronounced minimum with an obvious
$\gamma$-deformation in the PES for the $0^{+}$ state, which is in
disagreement with the results of ref.\cite{Bender08}, where a
calculation with 3DAMP+PNP based on non-relativistic Skyrme HFB
shows a pronounced triaxial minimum with $\beta=0.6$ and
$\gamma\approx16^\circ$ for the $0^{+}$ state. The energy gain due
to the restoration of rotational symmetry in the ground state of
$^{24}$Mg in the present investigation is about $5$~MeV. The minimum
on the PES for the $0^+$ state has a larger deformation
$\beta_p=0.56$ close to the data $\beta=0.57$~\cite{Fewell79}.
Keeping in mind the strong pairing gaps found in our mean-field
calculations, the introduction of PNP in our calculations is not
expected to introduce great changes in the results.

 \begin{figure*}[]
  \includegraphics[width=7cm]{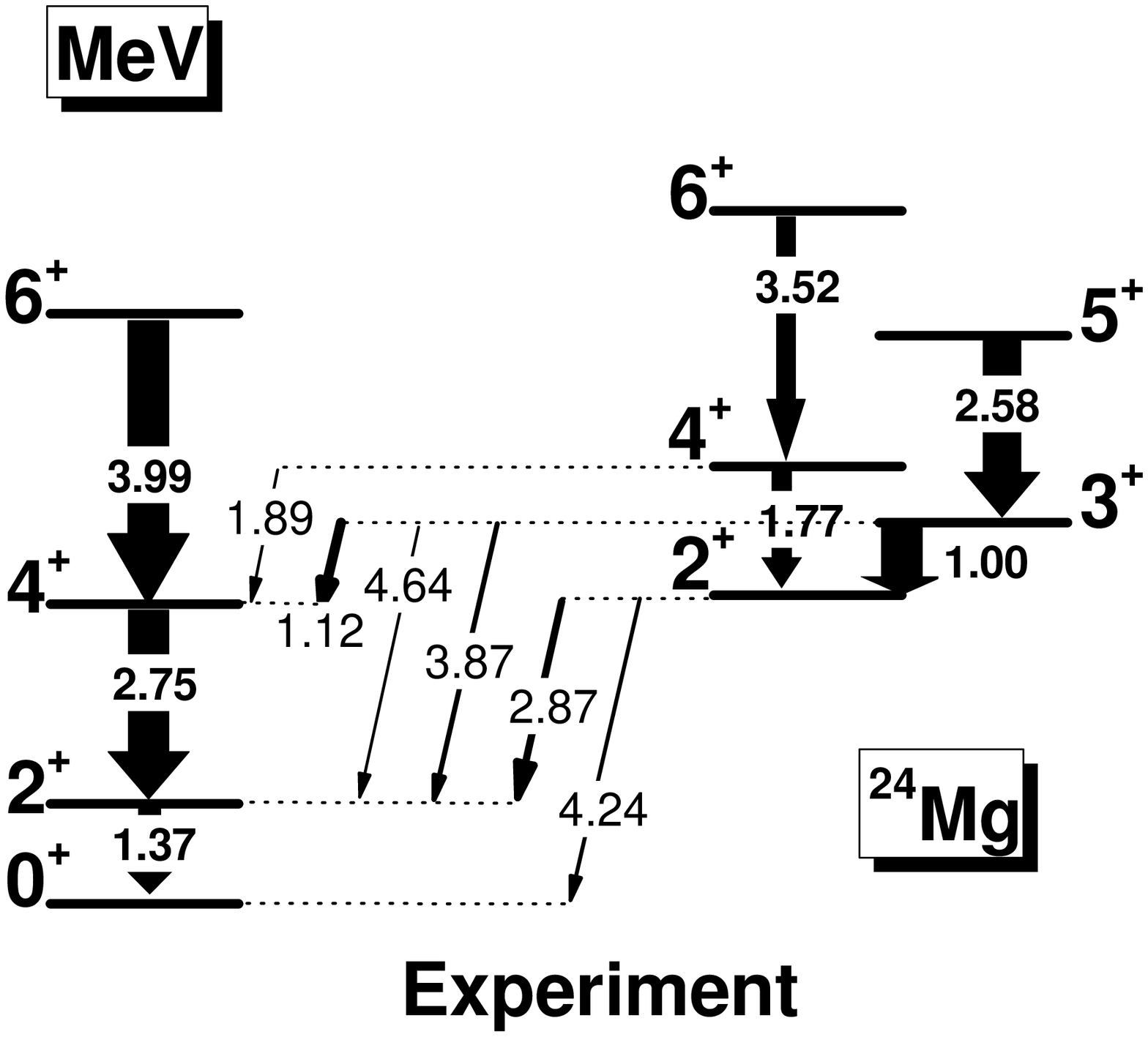}
  \includegraphics[width=7cm]{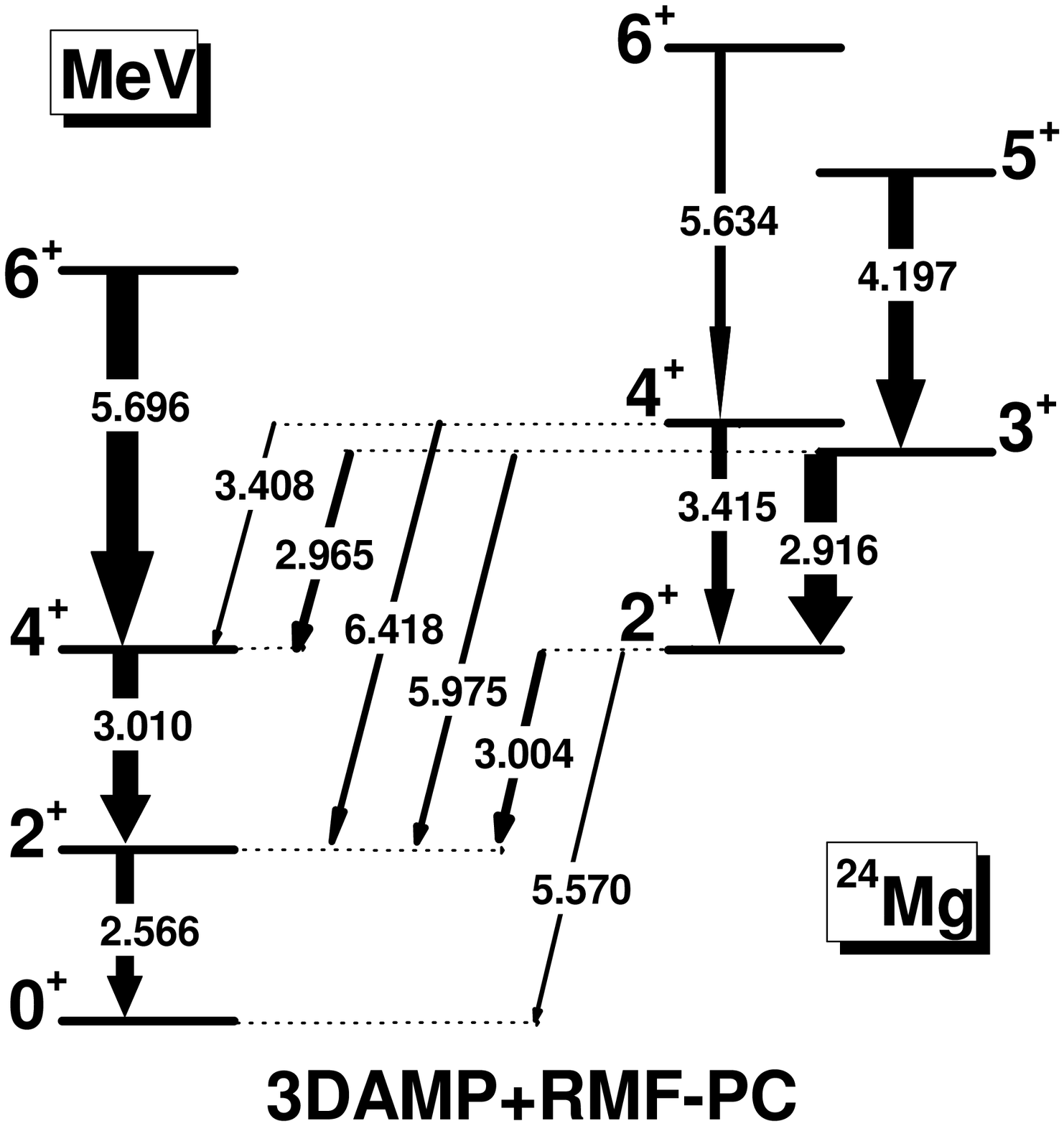}
   \caption{Comparison between the theoretical and experimental
   bands in $^{24}$Mg. The excitation energies are
   normalized to the ground state and given in MeV. The widths of
   the arrows represent the $B(E2\downarrow)$ values.
   The experimental data is taken from Refs.~\cite{Branford75} and \cite{Keinonen89}.}
   \label{fig3}
 \end{figure*}

The right panel of Fig.~\ref{fig2} shows that there is a minimum in
the PES for the $2^{+}$ state localized at
$\beta\approx0.55,~\gamma\approx10^\circ$. Using the 3DAMP+RMF-PC
approach we calculate the spectrum of $^{24}$Mg based on this
triaxial mean-field state. This is the only way to obtain $K=2$
bands in our calculation. In Fig.~\ref{fig3} we show this spectrum
and compare it with the known experimental data. The observed
energies are qualitatively reproduced by the theory, even though the
calculated levels are systematically higher in energy, which might
be due to the fact that all the excited states are projected from
the same mean-field state. The projected energy spectrum would be
much more compressed with the inclusion of a cranking term in the
mean-field calculation~\cite{Hara82,Baye84,Zdunczuk07}. This is a
strong indication that, for a quantitative understanding of the
spectra, the variation should be carried out after the projection.
Of course this goes beyond the present computer capabilities.

In conclusion, we have studied the properties of the ground state
and low-lying excited states in $^{24}$Mg using a full three
dimensional angular momentum projection on top of a relativistic
mean-field calculation based on a point-coupling Lagrangian
(3DAMP+RMF-PC) with monopole pairing. Ground state properties are
found to be reproduced rather well with this approach, even though
no pronounced minimum with obvious triaxial deformation is present.
A minimum with $\beta\approx0.55,~\gamma\approx10^\circ$ has been
found on the PES of the first $2^{+}$ state, on top of which the
experimentally observed excitation energies and the B(E2) transition
probabilities can be qualitatively reproduced. However, the
predicted spacing between the levels is overestimated by this
approach. It is expected that this can be cured by the introduction
of a cranking term at the mean-field level. These results hint that
a full GCM approach is needed for a full understanding of the
properties of low-lying states in $^{24}$Mg, and work towards this
goal is in progress.

 \bigskip
 \leftline{\bf ACKNOWLEDGMENTS}
 Helpful discussions with D. Vretenar are gratefully acknowledged.
This research has been supported by the Asia-Europe Link Project
[CN/ASIA-LINK/008 (094-791)] of the European Commission, the
National Natural Science Foundation of China under Grant No.
10775004, 10221003, 10720003, 10705004, the Bundesministerium
f\"{u}r Bildung und Forschung, Germany under project 06 MT 246 and
by the DFG cluster of excellence \textquotedblleft Origin and
Structure of the Universe\textquotedblright\
(www.universe-cluster.de).

\vspace{0.3cm}
 

\end{document}